# Photoacoustic digital tooth and image reconstruction of tooth root

Yuting Shen[1], Yiyun Wang[1], Chengxiao Liu[2], Niansong Ye[3], Feng Gao[1], Lunguo Xia[2,*], Bing Fang[2,*], Fei Gao[1,*]

[1] *ShanghaiTech University, Shanghai, China*
[2] *Shanghai Ninth People's Hospital, Shanghai, China*
[3] *Shanghai Huaguang Dental Clinic, Shanghai, China*
* *Corresponding authors: xavierxlg@126.com, fangbing@sjtu.edu.cn, gaofei@shanghaitech.edu.cn*

*Abstract*—**Imaging of teeth is very important to doctors in the diagnosis and treatment of dental diseases. The main imaging modality currently used is Cone-Beam Computed Tomography (CBCT), which however suffers from ionizing radiation causing potential damage to human body. In this work, photoacoustic imaging is proposed for the imaging of tooth, specifically the tooth root. A photoacoustic digital phantom of the tooth is generated based on clinical CBCT data. The roots encased by alveolar bone are imaged by using the realistic photoacoustic digital phantom in the simulation study. Several image reconstruction algorithms are used and compared to remove the artifacts caused by heterogeneous acoustic velocity distribution.**

*Keywords—photoacoustic imaging, numerical phantom, tooth imaging*

## I. Introduction

In the diagnosis and treatment of dental diseases, accurate tooth imaging and segmentation are very important [1]. For example, in root canal therapy, the clear evaluation of the tooth root boundary is the very important first step [2]. In orthodontic process, the segmentation of alveolar bone and root is also a key step [3]. Reconstruction activities such as absorption and hyperplasia occur on the surface of the root leading to the destruction of the periodontal membrane and gingival tissue, which may result in alveolar bone atrophy. It would further increase the difficulty of tooth movement. In clinical practice, the effect of orthodontic operation usually depends on the experience of the doctor.

The current golden-standard tooth imaging modality is oral and maxillofacial cone beam CT (CBCT) [4], which reconstruct the tooth through three-dimensional cone beam X-ray scan inside the mouth. However, CBCT is an imaging method with radiation, which would cause harm to the human body. Besides, the contrast between teeth and roots is low, so the root boundary will be blurred in the segmentation. Another way to image the tooth is high-frequency ultrasound (HFUS) imaging [5]. It is a non-invasive imaging modality that can image hard tissues behind soft tissues. However, its clinical application is still relatively limited due to issues such as scanner size, scanning time, imaging accuracy and the cost.

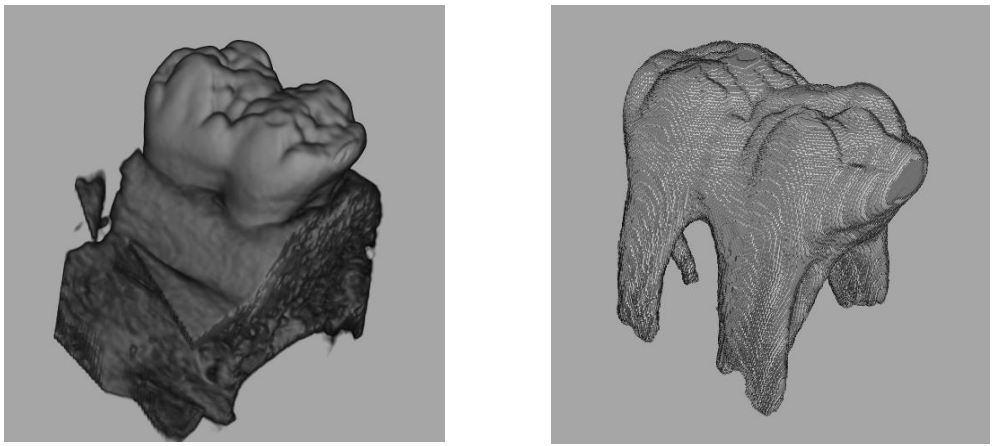

Fig. 1. The tooth phantom after segmentation.

Photoacoustic imaging is a new imaging method, which is non-ionizing and non-invasive [6] based on photoacoustic effect [7]. When exposed to laser, the biological tissue emits ultrasound signals following thermoplastic expansion, which can be detected by ultrasound transducers. It combines the high specificity of optical imaging with the deep penetration of ultrasonic wave. In this work, we propose to apply photoacoustic imaging to the imaging of teeth root. In the literature, the feasibility of photoacoustic teeth imaging was verified with *in vitro* experiments by Schneider et al [8]. Sampathkumar et al [9] proposed to image the early-stage dental caries with all-optical photoacoustic imaging and achieve μm level resolution. In our work, the feasibility of photoacoustic imaging for tooth root is verified through digital phantom generation and simulation experiments. As for the artifact caused by the heterogeneous acoustic velocity, multiple image reconstruction algorithms are applied and compared.

## II. METHODS

### A. Digital tooth phantom generation

In this section, the process of generating the digital tooth phantoms is introduced. In order to generate a numerical phantom that is as realistic as possible, CBCT data of pig tooth are utilized. We took two of the teeth as our phantom and cut them out from the CBCT volume data. Since the purpose of photoacoustic imaging is to distinguish the number of tooth root, one of the phantom is canine with single tooth root and the other one is premolar with 6 tooth roots. The cropped CBCT data is segmented into 3 layers, including gingiva, alveolar bone and tooth root , according to the physical value. One example is shown in Fig. 1.

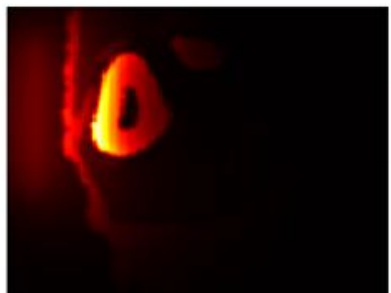

(a) The simulation result of light absorption, which is used as initial pressure distribution.

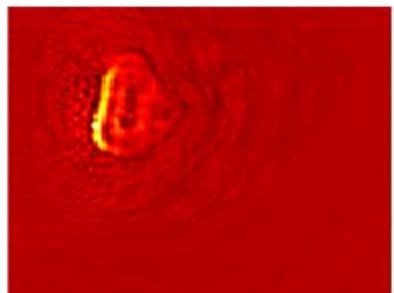

(b) The reconstruction result of time reversal.

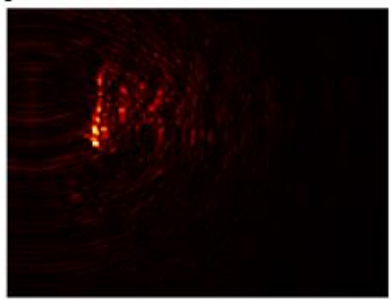

(c) The reconstruction result of DAS.

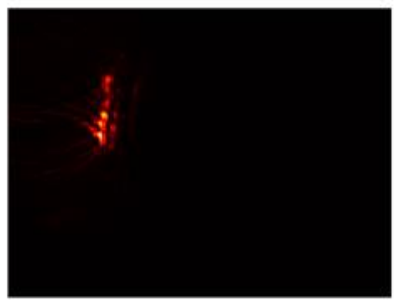

(d) The reconstruction result with cut signal.

Fig. 2. Reconstruction results of single tooth root simulation.

To make the segmented phantom smooth and connected, double threshold detection algorithm is used. Take root layer as an example, a maximum value and a minimum value are set for the segmentation. Pixels above the maximum value are directly put into the root category. Pixels below the maximum value but above the minimum value are also selected if they are adjacent to the selected root pixels. In this way, CBCT volume data are labeled as 3 types of structures.

*B. Optical simulation*

In this section, we introduce the process of optical simulation. In photoacoustic effect, biological tissues absorb the energy of the laser and generate heat to produce acoustic waves. There are several parameters in optical simulation, including optical absorption coefficient $\mu_a$, scattering coefficient $\mu_s$ and refractive index g. The wavelength is set to 1064 nm. Since there are few studies on the optical parameters of teeth, we use skull's parameters to simulate the alveolar bone and use soft tissue's parameters to simulate gingiva. All the optical parameters are found in related works [10]-[13] and shown in Table I. The physical parameters are found in IT'IS Foundation: https://itis.swiss/virtual-population/tissue-properties/database/.

TABLE I. OPTICAL AND ACOUSTIC PROPERTIES OF THE TOOTH PHANTOM

|  | $\mu_a$ (cm$^{-1}$) | $\mu_s$ (cm$^{-1}$) | g | $\rho$ (kg/m$^3$) | c (m/s) |
|---|---|---|---|---|---|
| Gingiva | 0.5 | 97 | 1.335 | 1102 | 1620.7 |
| Alveolar bone | 0.225 | 14.25 | 1.560 | 1900 | 2117.5 |
| Dentin | 3 | 260 | 1.540 | 2200 | 3600 |

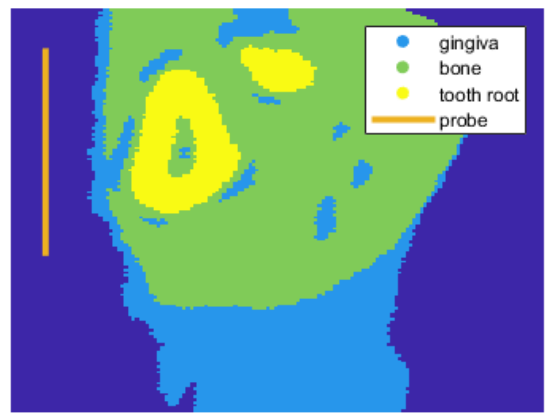

Fig. 3. The cross section of photoacoustic digital tooth phantom with single root. The yellow line show the positions of the linear probe array.

The optical simulation process is carried out through NIRFAST, which is a toolbox in MATLAB [14]-[16]. NIRFAST can model multi-wavelength light propagation through tissues of the given phantom. In our case, it is used to generate the light absorption distribution, which is the product of light flux and light absorption coefficient. Since we are imaging the cross sections of the tooth, a 2-D simulation is carried out. The phantom is a cross section taken from the volume data and an example is shown in Fig. 2. To use NIRFAST, a mesh grid is first generated. The optical parameters are set to different components in the phantom. The position of the light source are allocated at the left edge, which is the same position as the probes.

*C. Acoustic simulation*

After the light absorption distribution is generated, the acoustic simulation is carried out with k-Wave toolbox in MATLAB [17]. Since the photoacoustic initial pressure distribution is considered to be proportional to the light absorption, optical absorption distribution obtained from optical simulation is directly used as the initial pressure distribution. The density distribution map is generated according to the photoacoustic digital phantom's physical properties. For the background and gingiva, 1% variance of acoustic velocity is added. For alveolar bone and dentin, 2% variance of acoustic velocity is added.

*D. Image reconstruction algorithms*

In this section, the process of signal processing and image reconstruction are introduced. The signal from the gingiva is removed from the time-domain signal. To find the appropriate position of the cutting point, simulation of ultrasound imaging is performed to find the depth of gingiva. The image is first reconstructed with Delay and Sum (DAS). Since the speed of sound in DAS algorithm is uniform by default, the reconstructed images could be distorted and have artefacts. In addition to DAS, time reversal (TR) algorithm is also performed. In time-reversal reconstruction, the speed of sound distribution is known, which shows the best image reconstruction results.

To address the distortion caused by the heterogeneous speed of sound in DAS algorithm, one solution is to remove the corresponding part of sound velocity distortion, and truncate the original data in time domain [18],[19]. In the case of teeth imaging, ultrasound signal can help identify the interface between gum and alveolar bone, but it is difficult to distinguish the alveolar bone from the tooth root. We set a cutting point of about twice the flight time of the probe to the alveolar bone, and the signals received after the cutting point are removed. On the one hand, the length is long enough to preserve the signals from the root. On the other hand, the signal is kept short as much as possible to reduce the influence from the reflection of the gingiva. Besides, according to the position of alveolar bone, the speed of sound are set to two different values, 1600 m/s and 2100 m/s, in the reconstruction, corresponding to gingiva layer and alveolar bone.

III. SIMULATION EXPERIMENTS AND RESULTS

*A. Case 1: Single root digital phantom*

A cross section is selected from the single root photoacoustic digital phantom. The size of the phantom is 146×146 pixels, and the grid size is 0.125 mm. A linear array of 32 transducers is placed along the edge shown in Fig. 2. The central frequency is set to 7.5 MHz and the bandwidth is set to 80%. The position of light source is the same as the transducers. The simulation result of initial pressure distribution is shown in Fig. 3a, and the TR reconstruction result is shown in Fig. 3b. The reconstruction result by DAS is shown in Fig. 3c. The reconstruction result of using cut signals is shown in Fig. 3d. Since the heterogeneous distribution of sound speed in TR is known, the reconstruction result is the most accurate one regarding the location of the root. DAS algorithm can just forms a blurred surface, and there are artifacts behind the surface. After the signal is cut properly, the artifacts are removed and the surface is more smooth with higher contrast.

## B. Case 2: Multiple roots digital phantom

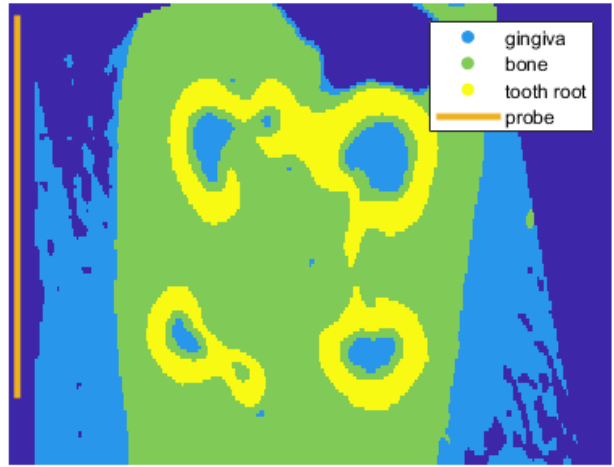

Fig. 4. The cross section of tooth phantom with multiple roots. The yellow line show the positions of the linear probe array.

A cross section is selected from the multiple root photoacoustic digital phantom shown in Fig. 4. The size of the phantom is $135 \times 251$ pixels, and other settings are the same as the one in single root phantom. The simulation result of initial pressure distribution is shown in Fig. 5a and the TR reconstruction result is shown in Fig. 5b. The reconstruction result by DAS is shown in Fig. 5c, and the reconstruction result after cutting is shown in Fig. 5d. It can be found that the gingiva layer is much thicker than the single root case, which leads to a strong light absorption in gingiva instead of the root. After removing the signal from the gingiva, TR could reconstruct the surface of the front two roots. In DAS result, the reflection from gingiva strongly affects the reconstruction, which also appears in TR result as a line behind the root surface, and the intensity of root surface is too low to be found. After the photoacoustic signal is cut properly, the position of the root is corrected and the reflection artifacts are

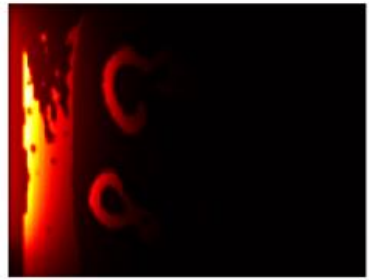

(a) The simulation result of light absorption, which is used as initial pressure distribution.

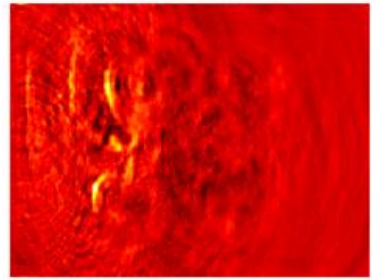

(b) The reconstruction result of time reversal.

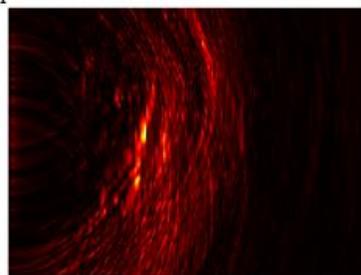

(c) The reconstruction result of DAS.

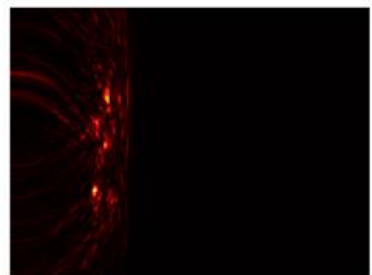

(d) The reconstruction result with cut signal.

Fig. 5 Reconstruction results of multiple roots simulation.

removed. Although the front surface is not as clear as TR one, it can be found at a correct position. Some arcing artifacts exist on the result may be due to the fact that the cutting point only considers the transverse distance, so when the signal is used for the reconstruction in the oblique direction, it would produce artifacts.

A quantitative comparison is provided in the table below to compare the performance of this two algorithms in reconstructing the tooth root. Peak signal-to-noise ratio (PSNR) and Structural Similarity (SSIM) are calculated based on the initial distribution. Images are normalized to eliminate the differences in absolute values. We can find that after cutting the signal, DAS gets similar results as TR. Both SSIM are not high enough, which may be caused by the artifacts.

TABLE II. QUANTITATIVE INDEX OF THE RECONSTRUCTION RESULT FROM TR AND DAS WITH CUT SIGNAL

|  |  | PSNR | SSIM |
|---|---|---|---|
| Single root | TR | 24.15 | 0.69 |
|  | DAS with cut signal | 21.86 | 0.71 |
| Multiple root | TR | 22.46 | 0.74 |
|  | DAS with cut signal | 22.37 | 0.75 |

## IV. CONCLUSION

In this work, photoacoustic digital tooth phantoms are built from CBCT data. They are used for photoacoustic imaging to verify the feasibility of distinguishing root and alveolar bone. Both optical simulation and acoustic simulation are carried out. Several image reconstruction algorithms are performed to treat the artifacts caused by heterogeneous sound speed distribution.

The simulation results show that photoacoustic imaging of root is feasible, since the light absorption coefficient of the root is much higher than that of the alveolar bone. One challenge is the distortion caused by non-uniform sound speed distribution. This could be solved by more accurate modeling of the bone or iterative optimization-based methods. In the future work, ex vivo experimental validation would be performed to further validate and improve the photoacoustic imaging capability of tooth root.

## ACKNOWLEDGMENT

This research was funded by National Natural Science Foundation of China (61805139), United Imaging Intelligence (2019X0203-501-02), Shanghai Clinical Research and Trial Center (2022A0305-418-02), and Double First-Class Initiative Fund of ShanghaiTech University (2022X0203-904-04).